\documentstyle[12pt,epsf]{article}
\newcommand{\newsection}{    % Numeration of eqs. is automatic
\setcounter{equation}{0}
\section}
\def\appendix#1{
\addtocounter{section}{1}
\setcounter{equation}{0}
\renewcommand{\thesection}{\Alph{section}}
\section*{Appendix \thesection\protect\indent #1}
\addcontentsline{toc}{section}{Appendix \thesection\ \ \ #1}
}
\newcommand{\rf}[1]{(\ref{#1})}

\def\be{\begin{equation}}
\def\ee{\end{equation}}
\newcommand{\beq}{\begin{equation}}
\newcommand{\eeq}{\end{equation}}
\newcommand{\bea}{\begin{eqnarray}}
\newcommand{\eea}{\end{eqnarray}}

\newcommand{\om}{\omega}

\newcommand{\k}{\kappa}

\newcommand{\non}{\nonumber}

\newcommand{\psid}{\psi^{\dagger}}
\newcommand{\phid}{\phi^{\dagger}}

\begin{document}
\topmargin 0pt
\oddsidemargin 5mm
\headheight 0pt
\headsep 0pt
\topskip 9mm

\hfill NBI-HE-97-17 

\hfill TIT/HEP-353

\addtolength{\baselineskip}{0.20\baselineskip}
\begin{center}
\vspace{26pt}
{\large \bf {The two-point function of $c=-2$ matter \\
             coupled to 2D quantum gravity}}
\end{center}

 \vspace{26pt}
\begin{center}
{\sl J.\ Ambj\o rn, C.\ Kristjansen}\\                                      
\vspace{6pt} 
The Niels Bohr Institute \\ 
Blegdamsvej 17,
DK-2100 Copenhagen \O, Denmark\\
\vspace{18pt}

and
 
\vspace{18pt}
 {\sl Y.\ Watabiki}\\ 
 \vspace{6pt}
 Tokyo Institute of Technology \\
 Oh-okayama, Meguro,
 Tokyo 152, Japan \\
 \end{center}
\vspace{20pt} 
\begin{center}
{\bf Abstract}
\end{center}

We construct a reparametrization invariant two-point function for $c=-2$ 
conformal matter coupled to two-dimensional quantum gravity. From the
two-point function we extract the critical indices $\nu$ and
$\eta$. The results support the quantum gravity version of Fisher's
scaling relation. Our approach is based on the transfer matrix
formalism and exploits the formulation of the $c=-2$ string as an
$O(n)$ model on a random lattice.

\noindent
\newpage

\newsection{Introduction \label{intro} }
Two-dimensional gravity has been a very useful laboratory for the 
study of interaction between matter and geometry. In particular, the 
so-called transfer matrix formalism \cite{KKMW93,Wat95} has provided us 
with a new tool to analyse {\it quantum geometry}. It allows us to 
study the fractal structure of space time and using this formalism 
it has been possible to calculate a reparametrization invariant two-point 
function of pure gravity \cite{AW95}. This two-point function has a
number of nice properties \cite{AW95}: 
\begin{enumerate}
\item Both the short distance and the long 
distance behaviour of the two-point point function reflect directly the 
fractal structure of quantum space-time.
\item The two-point function makes possible  a definition of a mass gap
in two-dimensional quantum gravity. 
\item In a regularised theory (e.g.\ two-dimensional quantum gravity 
defined by means of dynamical triangulations) this mass gap plays 
the same role as in the theory of critical phenomena: it monitors
the approach to the critical point.
\item From the two-point function
it is possible to define the same critical exponents as in the 
theory of critical phenomena, i.e.\  the mass gap exponent $\nu$,
the anomalous scaling exponent $\eta$ and the susceptibility exponent 
$\gamma_{\rm str}$.
\item Although the exponents take unusual values from the point of 
view of conventional field theory ($\eta=4$, $\gamma_{\rm str}=-1/2$), 
they nevertheless satisfy Fisher's scaling relation 
$\gamma_{\rm str} = \nu (2-\eta)$.
%dh%\item The fractal dimension $d_h$ of space-time is related to 
%dh%the critical exponent $\nu$ by $d_h = d_H \equiv 1/\nu$.
\end{enumerate}
More specifically,
 the renormalised two-point function of pure two-dimensional 
quantum gravity reads 
\begin{equation}
G_\Lambda (D) = \Lambda^{3/4} \, 
\frac{\cosh (\sqrt[4]{\Lambda} \, D)}
{\sinh^3( \sqrt[4]{\Lambda}\, D)},
\label{2point}
\end{equation}
where $D$ is the geodesic distance between two marked points. 
It is readily seen that the two-point function (\ref{2point})
(and its discretised version)
satisfy the points 1.--5.\ \cite{AW95}. 
We expect 1.--5.\ to be valid 
even if matter is coupled to quantum gravity, although with different 
critical exponents, and 1.--5.\ constitute the foundation of the successful
application of finite size scaling in two-dimensional quantum gravity
\cite{CTBJ95,AJW95,AAMT96}. 

A major puzzle remains in two-dimensional quantum gravity. 
Formal constructions of a string field Hamiltonian for a $(m,m+1)$ conformal 
field theory coupled quantum gravity  suggest that \cite{IK94,IIKMNS94}
\begin{equation}
d_H \equiv \frac{1}{\nu} = 2m.
\label{haus1}
\end{equation}
This relation can also be written as 
\begin{equation}
d_H \equiv \frac{1}{\nu} = - \, \frac{2}{\gamma_{\rm str}}.
\label{haus2}
\end{equation}
However, this relation has not been observed in the numerical 
simulations \cite{CTBJ95,AJW95}. A number of problems have so far 
blocked decisive comparison between the theoretical prediction \rf{haus2}
and the numerical simulations. The (internal) Hausdorff dimension \rf{haus1}
is so large that the systems used in the numerical simulations 
might not have detected it. Secondly, additional assumptions about the 
field content go into the formal derivation of \rf{haus1} and 
in particular, the identification of the proper time $T$ of the string 
field Hamiltonian as the geodesic distance  becomes questionable. We
shall return to this point later, see also reference~\cite{AW96}.  

In this paper we determine the two-point function in the case where
$c=-2$ matter is coupled to quantum gravity. 
 As shown by David~\cite{Dav85}, the discrete version of the
$c=-2$ string~\cite{KKM85} can be 
mapped onto a zero-dimensional field theory via Parisi-Sourlas dimensional 
reduction~\cite{PS76}. This zero-dimensional field theory can be viewed
as a special version of the $O(n)$ model on a random
lattice~\cite{Kos89}; namely one for which $n=-2$. Our construction of
the two-point function for the system of
$c=-2$ matter coupled to quantum gravity
will be based on this equivalence with the $O(n)$ model. 
We start from the discretised
version of the model and construct the transfer matrix in the spirit 
of~\cite{IK94,IIKMNS94}, but without any additional assumptions on the
matter fields. The simplicity of the model allows us to determine the two-point
function exactly. We extract the critical indices $\nu$ and $\eta$, verify
that Fisher's scaling relation is fulfilled and find that our results 
support the relation~\rf{haus2}.
At the same time the $c=-2$ string
is a perfect model for numerical simulations and computer 
simulations allow a determination  of 
$\nu$ %dh% and $d_h$ 
with high precision.

The rest of this paper is organised as follows: in section~\ref{loopgas} 
we construct 
a string field theory for a general loop gas model which contains the
$O(n)$ model on a random lattice as a special case. Section~\ref{O(n)} is
devoted to the study of the $O(n)$ model itself and in section~\ref{n=-2} we
specialise to the case $n=-2$. Section~\ref{c=-2} 
contains a
detailed analysis of the $c=-2$ string and
finally in section~\ref{discuss} we discuss the exact results obtained 
and compare with numerical results. We also comment on the
implications of our results for the series of minimal unitary models
coupled to quantum gravity.

\newsection{Dynamical triangulations with coloured loops \label{loopgas}}
The $O(n)$ model on a random lattice is an example of a so-called loop
gas model. Loop gas models can be defined in very general 
settings (see for instance~\cite{Kon96} and references therein)
 and
play an important role in field theory as well as in statistical
mechanics. Hence, although our final aim is to study
the $O(n)$ model on a random lattice, 
we shall start out with a more general loop gas model.

\subsection{The model \label{model} }
We consider the set of two-dimensional closed connected complexes
obtained by gluing together $i$-gons ($i\geq3$) along pairs of
links. On these complexes we introduce coloured loops. The loops live
on the dual complex and their links connect centres of neighbouring
triangles belonging to the original complex. 
We restrict the loops to be closed,
self-avoiding and non-intersecting and we assume that they come in
$N_c$ different colours. We denote those triangles (links) in the
complex which are traversed by loops as decorated triangles (links)
and those not traversed by loops as non-decorated triangles (links).
All $i$-gons with $i\geq 4$ are per definition non-decorated.
We define the partition function of our model by
\beq
Z=\sum_{T \in {\cal T}} g_s^{h-1}\sum_{\{{\cal L}\}}
\frac{1}{C_T(\{ {\cal L}\})} \prod_{i\geq 3}g_i^{N_i}
\prod_{a=1}^{N_c}\kappa_a^{l_a}\,\lambda_a^{n_a}
\label{part}
\eeq
where ${\cal T}$ denotes the class of complexes described above, $T$ is a
complex in ${\cal T}$ and $\{ {\cal L}\}$ is a given loop configuration on
$T$ obeying the above given rules. The quantity $C_T(\{ {\cal L}\})$
is the order of the automorphism group of $T$ with the loop
configuration $\{ {\cal L}\}$. The quantity $h$ is the genus of the
complex, $N_i$ the number of non-decorated $i$-gons and $n_a$ the number
of loops of colour $a$. Finally $l_a$ is the total length of loops with
colour $a$ which is equal to the number of decorated triangles
carrying the colour $a$.

\subsection{The string field theory \label{sft} }

\begin{figure}[hbt]
\vspace{1.0cm}
\centerline{\epsfbox{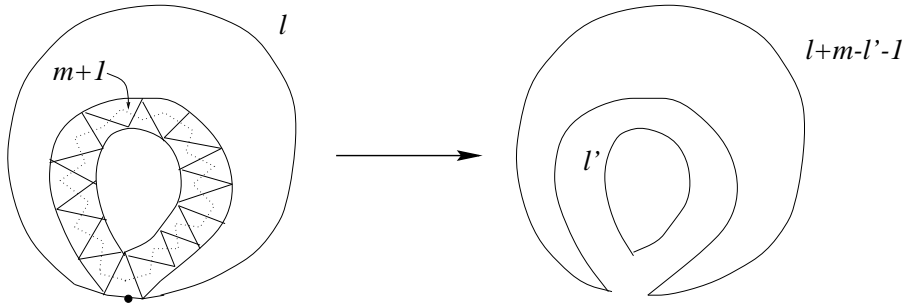}}
\caption[ppp]{The minimal step decomposition in the case where the marked
link belongs to a decorated triangle. The dotted line illustrates the loop.}
\label{deform}
\end{figure}

We shall now, corresponding to the model~\rf{part}, write down a string
field theory for strings consisting of only non-decorated links. To do
so it is necessary to define  a distance or a time variable on the
complexes introduced above.
 There exist two different approaches to this problem, known as
the slicing decomposition~\cite{KKMW93} and the peeling
decomposition~\cite{Wat95}. As it will become clear shortly the
presence of the loops on the surfaces makes it an advantage to use
 the peeling
decomposition. Let us consider a disk with a boundary consisting of
$l$ non-decorated links, one of which is marked. Our minimal step
decomposition will take place at the marked link. To describe the
deformation we introduce string fields, $\psid(l)$ and
$\psi(l)$ which respectively creates and annihilates a
closed string of non-decorated links having length $l$ and one marked
link. The string fields obey the following commutation relations, 
\bea
\left[\psi(l),\psid(l')\right]&=&\delta_{l,l'} \, , \label{com1}\\
\left[\psid(l),\psid(l')\right]&=&\left[\psi(l),\psi(l')\right]=0 \, .
\label{com2}
\eea
Expressed in terms of the string fields we define
our minimal step decomposition 
by~\footnote{Here it is understood that $\psid(l=0)=1$ and
$\delta \psid(l=0)=0$.}
\bea
\delta\psid(l)&=&-\psid(l)+
\sum_{i=3}^{\infty}g_i\psid(l+i-2)+\sum_{l'=0}^{l-2}\psid(l')\psid(l-l'-2)
\nonumber\\
&& + \sum_{a=1}^{N_c}\lambda_a\sum_{m=0}^{\infty}\kappa_a^{m+1}\sum_{l'=0}^m
\left(\!\!\begin{array}{c} m \\ l' \end{array}\!\!\right)
\psid(l')\psid(l+m-l'-1)\nonumber\\
&& +2g_s\sum_{l'=1}^{\infty}\psid(l+l'-2)l' \psi(l').
\label{delta}
\eea
In case the marked link does not belong to a decorated triangle
the minimal step decomposition of the surface is
defined exactly as in the pure
gravity case, i.e.\ the decomposition consists in removing either an
$i$-gon or a double link~\cite{Wat95}. The removal of an $i$-gon
always increases the length of the string by $i-2$. (As explained
in~\cite{Wat95} double links are supplied when necessary.) The removal
of a double link either results in the splitting of one string into
two or the merging of two strings into one. In the latter case a
handle is created. The creation of a handle is associated with a
factor of $2g_s$. This accounts for the terms in the
first and the third lines of~\rf{delta}. 
The term in the second line describes the
minimal step decomposition in the case where the marked link belongs
to a decorated triangle. This situation is illustrated in
figure~\ref{deform}. 
Let us assume that the loop which passes through
the triangle with the marked link has length $m+1$ ($m\geq0$). 
It hence passes through $m+1$ triangles. 
Of these $m+1$ decorated triangles a certain
number, say, $l'$ will have their base on the inner side of the loop
and the remaining $m+1-l'$ triangles will have their base on the outer
side of the loop. Our minimal step decomposition consists in removing
all $m+1$ triangles along the loop so that the initial string of length
$l$ is replaced by two new strings of length $l'$ and $l+m-1-l'$
respectively, see figure~\ref{deform}. 
A removed loop of colour $a$ is accompanied by a factor $\lambda_a$
and each removed triangle along the loop is accompanied by a factor
$\kappa_a$.
The factor 
$\left(\!\!\begin{array}{c} m \\ l' \end{array}\!\!\right)$ 
counts the number
of possible orientations of the triangles along the loop. 

We now perform a discrete Laplace transformation of our creation and
annihilation operators. The Laplace transformed versions of 
$\psi^{\dagger}(l)$ and $\psi(l)$ are defined by~\footnote{ 
Throughout this paper we will denote variables which refer to the loop
length as $l$, $l'$ and $m$ and the conjugate variables as $p$, $q$, $s$
and $s'$. In particular, for a given function or operator $f$, $f(l)$
will refer to its loop length version and $f(p)$ to its Laplace
transform as defined in equation~\rf{Laplace}.}
\beq
\psid(p)=\sum_{l=0}^{\infty} \frac{1}{p^{l+1}}\,\psid(l)
\hspace{0.5cm} (|p|> p_c),
\hspace{1.0cm} \psi(q)=\sum_{l=1}^{\infty}\frac{1}{q^{l+1}}\psi(l)
\hspace{0.5cm}(|q|>q_c),
\label{Laplace}
\eeq
and we assume that these expressions make
 sense for $|p|$ and $|q|$ larger than
some critical values $p_c$ and $q_c$ respectively. 
The inverse transformations to~\rf{Laplace}
can be written as
\beq
\psid(l)=\oint_{|s|=p_c}\frac{ds}{2\pi i} s^l \, \psid(s) ,
\hspace{1.0cm}\psi(l)=\oint_{|s|=q_c}\frac{ds}{2\pi i} s^l \, \psi(s) ,
\eeq
where the integrals can be evaluated by taking the residues at
infinity. Here and in the following we use the convention that 
unless otherwise indicated
contours are oriented counterclockwise. 
The Laplace transformed versions of the commutation relations~\rf{com1}
and~\rf{com2} are
\bea
\left[\psi(q),\psid(p)\right] &=&\frac{1}{pq(pq-1)}, \label{com1a} \\
\left[ \psid(p),\psid(p')\right] &=&\left[\psi(q),\psi(q')\right]=0. 
\label{com2a}  
\eea
We here need to require $p_c q_c = 1$. 
We next rewrite the string deformation equation~\rf{delta} 
in the Laplace transformed language. 
This is most easily done by applying the operator
$\sum_{l=1}^{\infty}\frac{1}{p^l}$
to both sides of~\rf{delta}. 
The treatment of the
terms in the first and third lines is standard while the treatment of
the term in the second line is less trivial. 
We have for $|p|>p_c$
\bea
I(p,\kappa)&\equiv & \sum_{l=1}^{\infty}\frac{1}{p^l}
\sum_{m=0}^{\infty}\k^{m+1}
\sum_{l'=0}^m \left(\!\!\begin{array}{c} m \\ l' \end{array}\!\!\right)
\psid(l')\psid(l+m-l'-1) \nonumber \\*
&=&\oint_{|s|=p_c}\frac{ds}{2\pi i}
\oint_{|s'|=p_c}\frac{ds'}{2\pi i}\,\frac{1}{p}
\sum_{l=0}^{\infty}\left(\frac{s}{p}\right)^l
\sum_{m=0}^{\infty} \k^{m+1}
\sum_{l'=0}^m \left(\!\!\begin{array}{c} m \\ l' \end{array}\!\!\right) 
s'^{l'} s^{m-l'} \psid(s)\psid(s')
\nonumber \\*
&=&\oint_{|s|=p_c}\frac{ds}{2\pi i}\oint_{|s'|=p_c}\frac{ds'}{2\pi i}
\frac{1}{p-s}\sum_{m=0}^{\infty}\k^{m+1}
\left(s'+s\right)^m \psid(s)\psid(s')
\nonumber \\*
&=&\oint_{|s|=p_c}\frac{ds}{2\pi i}\oint_{|s'|=p_c}\frac{ds'}{2\pi i}
\frac{1}{p-s}\,\frac{1}{\frac{1}{\k}-\left(s'+s\right)} \psid(s)\psid(s') .
\label{I}
\eea
Here the last equality sign is only valid if $|\k(s+s')|<1$. From this
we learn that we must require
\beq
p_c< \min_{a}\left|\frac{1}{2\kappa_a}\right|
\eeq
in order for the subsequent considerations to make sense.
\label{pc}
Stated
otherwise the model becomes singular as 
$p_c\rightarrow \min_{a}\left|\frac{1}{2\k_a}\right|$.
In equation~\rf{I} we can carry out the integration over $s'$. This
gives
\beq
I(p,\kappa)=\oint_{|s|=p_c}\frac{ds}{2\pi i}
\frac{1}{p-s}\psid(s)\psid(\frac{1}{\k}-s).
\label{Iint}
\eeq
We note that the product $\psid(s)\psid(\frac{1}{\kappa}-s)$ is
ill defined in both of the 
regions $|s|<p_c$ and $|\frac{1}{\kappa}-s|<p_c$. 
Due to this fact we can not immediately perform 
the second integration. 
We shall show later how to deal with this problem. 
For the moment we note that
we have
\bea
p\delta\psid(p)
&=&-\oint_{|s|=p_c}\frac{ds}{2\pi i}\frac{1}{p-s}V'(s)\psid(s)
   +\left(\psid(p)\right)^2+\sum_{a=1}^{N_c}\lambda_a\,I(p,\kappa_a) 
 \nonumber \\
&&+ 2 g_s \!\!\oint\limits_{|s|=p_c}\!\!\!\frac{ds}{2\pi i} 
  \frac{1}{p-s} \psid(s) 
  \frac{\partial}{\partial s} \left( \frac{1}{s} \psi(\frac{1}{s}) \right) 
\, , \label{deltap}
\eea
where we have made use of the following rewriting
\beq
-p\psid(p)+1+\sum_{i=3}^{\infty}g_ip^{i-1}
\left\{\psid(p)-\sum_{l=0}^{i-2}\frac{1}{p^{l+1}}\psid(l)\right\}
=-\oint_{|s|=p_c}\frac{ds}{2\pi i}\frac{1}{p-s}V'(s)\psid(s)
\eeq
with
\beq
V(s)=\frac{1}{2}s^2-\sum_{i=3}^{\infty}\frac{g_i}{i}s^i
\eeq
and $V'(s)=dV(s)/ds$.
We now introduce an operator ${\cal H}$, a Hamiltonian, which describes
the minimal step deformation of our surface or the time evolution of
the wave function. More precisely we define ${\cal H}$ by
\beq
l\delta\psid(l)=-\left[ {\cal H},\psid(l)\right],
\eeq
with the vacuum condition
\beq
{\cal H}|\mbox{vac}\rangle = 0 .
\eeq
It is easy to see that ${\cal H}$ 
can be expressed in the following
way
\bea
{\cal H}
&=&\sum_{l=1}^{\infty} l\delta\psid(l)\psi(l) 
= \oint_{|s|=p_c}\frac{ds}{2\pi i} 
  \delta\psid(s) s \frac{\partial}{\partial s}\left(
  \frac{1}{s} \psi \left(\frac{1}{s}\right)\right)  
\nonumber\\ 
&=&- \oint\limits_{|s|=p_c}\!\!\!\frac{ds}{2\pi i} 
\Biggl[ 
\left\{ - V'(s)\psid(s) + \left(\psid(s)\right)^2 
        + \sum_{a=1}^{N_c}\lambda_a\psid(s)
        \psid\left(\frac{1}{\k_a}-s\right)
\right\}
\frac{\partial}{\partial s}\left(\frac{1}{s}\psi\left(\frac{1}{s}\right)\right)
\nonumber \\
&&\hspace{2.5cm}
     + g_s \psid(s) 
       \left( \frac{\partial}{\partial s}
       \left( \frac{1}{s}\psi\left(\frac{1}{s}\right)\right)\right)^2 
\, \Biggr] \, . 
\label{H}
\eea
Here we have arrived at a simple form of the contribution coming from
the terms $I(p,\kappa_a)$ in $p\delta\psid(p)$ (cf.\ equation~\rf{deltap})
by changing the order of the two integrations. We note, 
however, that the simplification is only apparent. If we wanted 
to evaluate the contour integral in~\rf{H} explicitly we would 
again have 
to deal with the problem of the product $\psid(s)
\psid\left(\frac{1}{\kappa}-s\right)$ being ill defined in both of
the regions $|s|<p_c$ and $\left|\frac{1}{\kappa}-s\right|<p_c$.
%We stress that we have set $g_s=0$ in formulas~\rf{deltap} and~\rf{H}
%for simplicity only. The treatment of the term proportional to $g_s$ 
%poses no problems at this level. 
%We refer to reference~\cite{Wat95} for details.

\subsection{The transfer matrix \label{transfer1} }

\begin{figure}
\centerline{ \epsfbox{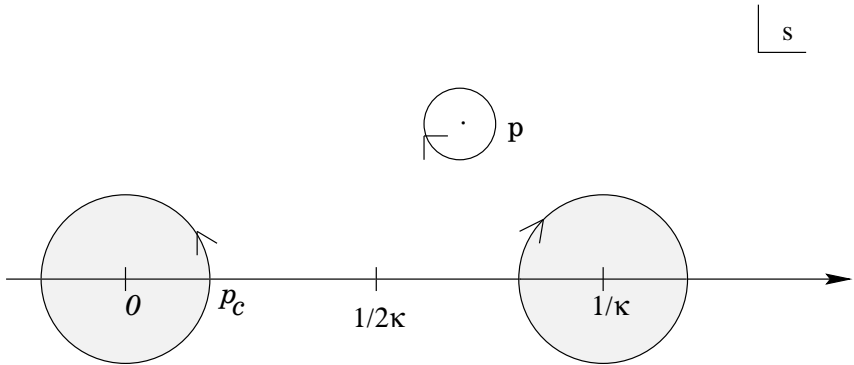}}
\caption[ppp]{Deformation of the contour of integration in integrals of the
type~\rf{I} and~\rf{J}. The shaded regions are forbidden regions for $p$.}
\label{contour}
\end{figure}

We shall now introduce a transfer matrix which describes the
propagation of a single string. For that purpose we need first 
to introduce the disk amplitude. The disk amplitude, $W(l)$, simply
counts the number of possible triangulations of the disk, constructed
in accordance with the rules of section~\ref{model}, with one boundary
component consisting of $l$ non-decorated links
one of which is marked. 
In the string field theory language it is given by
\beq
W(l)=\lim_{t\rightarrow \infty}
\langle \mbox{vac}|e^{-t{\cal H}_{\rm disk}}\psid(l)|\mbox{vac}\rangle, 
\eeq
where ${\cal H}_{\rm disk}\equiv \left.{\cal H}\right|_{g_s=0}$ 
and $t$ can be thought of as measuring the time evolving as the string
field propagates or the distance being covered as the surface is decomposed
(peeled).
Its Laplace transform, 
\beq
W(p)=\sum_{l=0}^{\infty}\frac{1}{p^{l+1}}\,W(l)
=\lim_{t\rightarrow \infty}
\langle \mbox{vac}|e^{- t{\cal H}_{\rm disk}} \psid(p)|\mbox{vac}\rangle, 
\eeq
is the so-called one-loop correlator.
Next, we define a modified
Hamiltonian, $\overline{{\cal H}}$, which generates the time evolution
of a single string
\beq
\overline{{\cal H}}=\sum_{l=1}^{\infty}\psid(l)
\left\{\left.\frac{\delta{\cal H}_{\rm disk}}{\delta\psid(l)}
\right|_{\psid(l)\rightarrow W(l)}\right\}.
\eeq
In the Laplace transformed picture we have
\bea
\lefteqn{\overline{{\cal H}} = 
\left. -\oint_{|s|=p_c}\frac{ds}{2\pi i}
\right\{-V'(s)\phid(s)+2W(s)\phid(s) \nonumber} \\
&&\hspace{0.6cm}\left.+\sum_{a=1}^{N_c}\lambda_a
\left[W\left(\frac{1}{\k_a}-s\right)\phid(s)+
W(s)\phid\left(\frac{1}{\k_a}-s\right)\right]\right\}
\frac{\partial}{\partial s}\left(\frac{1}{s}\psi\left(\frac{1}{s}\right)\right)
\eea
where
\beq
\phid(p)=\psid(p)-\frac{1}{p}
\eeq
i.e.\ we have excluded the trivial mode corresponding to a string of
zero length. We now define the transfer matrix for one-string
propagation by
\beq
G(p,q,t)=\langle \mbox{vac}\left|\psi(q)e^{-t\overline{{\cal H}}}
\phid(p)\right|\mbox{vac} \rangle.
\label{transferdef}
\eeq
 Differentiation with respect to $t$ gives
\beq
\frac{\partial }{\partial t}G(p,q,t)=-
\langle \mbox{vac}\left|\psi(q) e^{-t\overline{{\cal H}}}
\left[\overline{{\cal H}},\phid(p)\right]\right|\mbox{vac}\rangle.
\label{dT} 
\eeq
Using the commutation relations~\rf{com1a} and~\rf{com2a}
we find
\bea
\lefteqn{
\left[\overline{{\cal H}},\phid(p)\right]=
\left[\overline{{\cal H}},\psid(p)\right] \nonumber} \\
&=&
\frac{\partial}{\partial p}\left(
- \oint_{|s|=p_c}\frac{ds}{2\pi i}\frac{1}{p-s}V'(s)\phid(s)+
2W(p)\phid(p) + \sum_{a=1}^{N_c}\lambda_aJ(p,\kappa_a)\right)
\label{Hphi}
\eea
where
\beq
J(p,\kappa)=\oint_{|s|=p_c}
\frac{ds}{2\pi i}\,\frac{1}{p-s}\left\{W\left(\frac{1}{\k}-s\right)\phid(s)
+W(s)\phid\left(\frac{1}{\k}-s\right)\right\}\, .
\label{J}
\eeq
Inserting~\rf{Hphi} into~\rf{dT} we get
\bea
\frac{\partial}{\partial t}G(p,q,t)&=&-\frac{\partial}{\partial p}
\left(
-\oint_{|s|=p_c}\frac{ds}{2\pi i}\frac{1}{p-s}V'(s)G(s,q,t)
+2W(p)G(p,q,t)\right. \nonumber \\
&&+
\left.\sum_{a=1}^{N_c}\lambda_a
\langle \mbox{vac}\left|\psi(q)e^{-t\overline{{\cal H}}}J(p,\kappa_a)
\right| \mbox{vac} \rangle\right).
\label{dT2}
\eea
We see that the problem encountered in equation~\rf{Iint} and~\rf{H}
still persists. The integrand in~\rf{J} is ill defined in the region
$\left|\frac{1}{\kappa}-s\right|<p_c$ as well as in the region $|s|<p_c$.
 Let us now finally discuss how to deal with this problem.
First we note that the structure of the integrand in~\rf{Iint} and~\rf{J}
is similar. The integrand consists of a pole term and a factor which is
invariant under the change $s\rightarrow \frac{1}{\k}-s$. Such
integrals have an important symmetry.
To expose this symmetry, let us deform the
contour of integration into two new ones; one which encircles the
point $s=\frac{1}{\k}$ at the appropriate distance, given by
$|s-1/\k|=p_c$ and one which encircles the pole $s=p$,
see figure~\ref{contour}. 
Then we get for, say $J(p,\kappa)$
\beq
J(p,\kappa)=\phid(p)W\left(\frac{1}{\k}-p\right)
+\phid\left(\frac{1}{\k}-p\right)W(p)+\hat{J}(p,\kappa)
\label{J2}
\eeq
where we have simply picked up the residue at the pole $s=p$ and where
$\hat{J}(p,\kappa)$ is given by 
\beq
\hat{J}(p,\kappa)=-\oint_{|\frac{1}{\k}-s|=p_c}
\frac{ds}{2\pi i}\, \frac{1}{p-s} 
\left\{W\left(\frac{1}{\k}-s\right)\phid(s)
 + W(s)\phid\left(\frac{1}{\k}-s\right)\right\} .
\label{Jhat}
\eeq
Now performing the change of variable $s\rightarrow \frac{1}{\k}-s$
in~\rf{Jhat} one finds
\beq
\hat{J}(p,\kappa)=-J\left(\frac{1}{\k}-p,\kappa\right).
\label{JhatJ}
\eeq
It is easy to see that a similar symmetry is encoded in any integral
with the structure characterised above.
In this connection, let us add a comment on the
analyticity structure of equation~\rf{J2}. 
The first two terms on
the right hand side are well defined only if both the conditions $|p|>p_c$
and $\left|\frac{1}{\k}-p\right|>p_c$ are fulfilled while the third term
is well defined if $\left|\frac{1}{\k}-p\right|>p_c$. The sum of the three
terms, however, is well defined if $|p|>p_c$.
Exploiting the symmetry~\rf{JhatJ} 
one can, in certain special cases, by taking a  linear combination
of different versions of equation~\rf{dT2}, with suitably chosen values 
for the parameter $p$, eliminate the terms depending on $J(p,\k)$. In
the following section we shall see how this works for
the $O(n)$ model on a random lattice and leads to an exactly
solvable differential equation for the transfer matrix in the case $n=-2$.

\newsection{The $O(n)$ model on a random lattice \label{O(n)} }
The $O(n)$ model on a random lattice corresponds to the following
special case of the general loop gas model~\rf{part}
\beq
N_c=n,\hspace{0.7cm} \k_1=\ldots=\k_n=\kappa,
\hspace{0.7cm} \lambda_1=\ldots=\lambda_n=1
\label{On1}
\eeq
or equivalently
\beq
N_c=1,\hspace{0.7cm}\k_1=\k,\hspace{0.7cm}\lambda_1=n.
\label{On2}
\eeq

In this case the surface decomposition that we have presented 
in section~\ref{sft} is
equivalent to the one used in~\cite{Kos89II} to give a combinatorial
derivation of the Dyson-Schwinger equations for the $O(n)$ model on a
random lattice. Obviously these Dyson-Schwinger equations are
contained in the string field theory formulation. They can be
extracted as explained in~\cite{Wat95}. Furthermore, 
our string field deformation is similar in nature to the one
presented in~\cite{IK94} although in this reference several string
fields are introduced.
For the $O(n)$ model on a random lattice the 
existence of a singularity as $p_c\rightarrow \frac{1}{2\kappa}$ 
(cf.\ page~\pageref{pc})  is well-known and the analyticity
structure depicted in figure~\ref{contour} is completely equivalent to
the double cut nature of the saddle point equation encountered 
in the matrix model formulation of the model~\cite{Kos89,DK88,KS92,EZ92,EK95}.

In the following we shall make the restriction
\beq
g_4=g_5=\ldots=0.
\label{restrict}
\eeq
It is well known that when $n\in[-2,2]$ the $O(n)$ model on a random
lattice has a plethora of critical points at which the scaling behaviour can
be identified as that characteristic of conformal matter fields
coupled to two-dimensional quantum
gravity~\cite{Kos89,Kos89II,DK88,KS92,EZ92}. With a general potential
any minimal conformal model can be reached. With the
restriction~\rf{restrict}
still all minimal unitary models are within reach. We make this
restriction in order to obtain a simple contribution from the contour
integral term in the differential equation~\rf{dT2}. For the $O(n)$ model on a
random lattice given by the above choice of parameters,~\rf{dT2} reduces to
\beq
\frac{\partial}{\partial t}G(p,q,t)
=-\frac{\partial}{\partial p}\left(
\left[-V'(p)+2W(p)\right]G(p,q,t) + 
n\langle \mbox{vac} |\psi(q)e^{-t\overline{\cal H}}
J(p,\k)|\mbox{vac}\rangle
\right).
\eeq
Now,
inserting~\rf{J2} and~\rf{JhatJ} into~\rf{dT2} we arrive at
\bea
\lefteqn{\frac{\partial}{\partial t}G(p,q,t)=
-\frac{\partial}{\partial p}\left(\left[-V'(p)+2W(p)+
nW\left(\frac{1}{\k}-p\right)\right]G(p,q,t)\right. \nonumber} \\*
&&\left. +nW(p)G\left(\frac{1}{\k}-p,q,t\right)
-n\langle \mbox{vac}\Big|\psi(q)e^{-t\overline{{\cal H}}}J
\left(\frac{1}{\k}-p,\k\right)\Big|\mbox{vac}\rangle\right).
\label{dT3}
\eea
Hence we see that by subtracting from equation~\rf{dT3} the equation~\rf{dT2}
with $p$ replaced by $\frac{1}{\k}-p$ we can
eliminate the term involving $J\left(\frac{1}{\k}-p,\k\right)$.
The resulting equation
reads
\bea
\lefteqn{
\frac{\partial}{\partial t}
\left\{G(p,q,t)-G\left(\frac{1}{\k}-p,q,t\right)\right\}= 
\nonumber
} \\*
&&-\frac{\partial}{\partial p}
\left\{\left[-V'(p)+2W(p)+nW\left(\frac{1}{\k}-p\right)\right]
G(p,q,t)\right. \nonumber \\* 
&&+\left.
\left[-V'(\frac{1}{\k}-p)+2W\left(\frac{1}{\k}-p\right)
+nW(p)\right]G\left(\frac{1}{\k}-p,q,t\right)\right\}.
\label{dT4}
\eea
As we shall see in the following section for $n=-2$ and for a certain class
of potentials this equation contains enough information that the complete
transfer matrix can be extracted. We note that for $n\neq \pm 2$ we can
write
\bea
\lefteqn{\hspace{-1.0cm}
\frac{\partial}{\partial t}
\left\{G(p,q,t)-G\left(\frac{1}{\k}-p,q,t\right)\right\}= 
\nonumber} \\*
&&-\frac{\partial}{\partial p}
\left\{\left[2W_s(p)+nW_s\left(\frac{1}{\k}-p\right)\right]
G(p,q,t)\right. \nonumber \\* 
&&+\left.
\left[2W_s\left(\frac{1}{\k}-p\right)
+nW_s(p)\right]G\left(\frac{1}{\k}-p,q,t\right)\right\}
\label{dT5}
\eea
where $W_s(p)$ is the singular part of the disk amplitude, defined by
\beq
W(p) = W_r(p) + W_s(p)
\eeq
with
\beq
W_r(p) = \frac{2V'(p) - nV'(\frac{1}{\kappa}-p)}{4-n^2}.
\eeq

\section{The $O(-2)$ model \label{n=-2} }
\begin{figure}
\centerline{\epsfbox{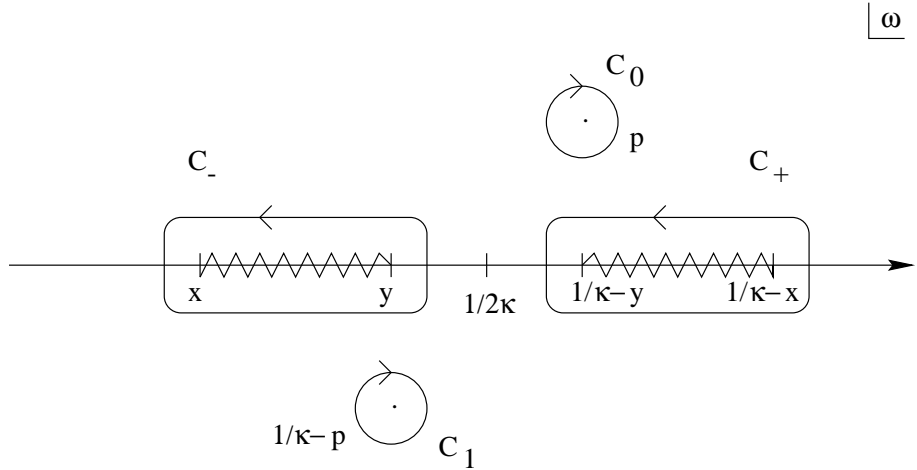}}
\caption[ppp]{Analyticity structure of $G_+(p)$ and $G_-(p)$ and the recipe
for going from one to the other (cf.\ eqn. \rf{G+-}). }
\label{int}
\end{figure}

Let us split the transfer matrix into two components in the
following way
\beq
G(p,q,t) = \frac{1}{2}\left\{G_+(p,q,t) + G_-(p,q,t)\right\}
\label{defG+-}
\eeq
where
\beq
G_{\pm}(p,q,t) = G(p,q,t)\pm G\left(\frac{1}{\k}-p,q,t\right).
\label{+-}
\eeq
The functions $G_+(p,q,t)$ and $G_-(p,q,t)$ are respectively symmetric
and antisymmetric under the transformation $p\rightarrow
\frac{1}{\k}-p$, i.e.
\beq
G_{\pm}(p,q,t) = \pm G_{\pm}(\frac{1}{\k}-p,q,t).
\label{parity}
\eeq
Now, let us restrict ourselves to considering an antisymmetric
potential, i.e.\ let us assume that
\beq
V'(p) = -V'\left(\frac{1}{\k}-p\right).
\label{V'}
\eeq
As for generic $n$, in order to obtain a simple contribution from the 
the contour integration term in the differential equation~\rf{dT2} we
furthermore restrict the degree of the potential. In this case it is 
sufficient to require
\beq
g_5=g_6=\ldots=0.\label{g5=0}
\eeq 
(We note that the relations~\rf{V'} and~\rf{g5=0} 
leave only one free parameter for the
potential.)
Then it appears that for $n=-2$ the differential equation~\rf{dT2} turns
into a closed equation for $G_-(p,q,t)$, namely
\beq
\frac{\partial}{\partial t}G_-(p,q,t)=
-\frac{\partial}{\partial p}
\left\{\left(-V'(p)+2W_-(p)\right)G_-(p,q,t)\right\}
\label{dTf}
\eeq
where in analogy with~\rf{+-} we have introduced
\beq
W_{\pm}(p)=W(p)\pm W\left(\frac{1}{\k}-p\right).
\eeq
As initial condition we have 
\beq
G_-(p,q,t=0)=\frac{1}{pq(pq-1)}-
\frac{1}{\left(\frac{1}{k}-p\right)q\left(\left(\frac{1}{\k}-p\right)q-1\right)}
.
\label{diffbound}
\eeq
The differential equation~\rf{dTf} allows us to determine the odd part of the
transfer matrix and as we shall show shortly
the analyticity structure of the problem allows us to
extract from $G_-(p,q,t)$ the function $G_+(p,q,t)$ and therefore
the full transfer matrix. (We note that the argument is not specific to 
the case $n=-2$ but $n=-2$ is the only case where we have a closed equation
for $G_-(p,q,t)$.)
The full transfer matrix
$G(p,q,t)$ is well defined for any $|p|>p_c$ while its two components
$G_+(p,q,t)$ and $G_-(p,q,t)$ are well defined only if we have both
$|p|>p_c$ and $|1/{\k}-p|>p_c$. It follows from the matrix model
calculations and equation~\rf{dTf} that the singularities of
$G(p)$ (where for simplicity
we have suppressed the dependence on $q$ and $t$)
 manifest themselves in the form of
cuts. Let us assume that $G(p)$ has only one cut $[x,y]$.
Then the functions $G_+(p)$ and $G_-(p)$ both have two cuts,
namely $[x,y]$ and $[\frac{1}{\k}-y,\frac{1}{\k}-x]$. 
To express $G_+(p,q,t)$ in terms of $G_-(p,q,t)$ we follow the line of
reasoning of reference~\cite{EK95}.
First, we express the fact that 
$G(p)$ is analytic along the interval $[1/\k-y,1/\k-x]$
\beq
G_+(p+i0) + G_-(p+i0) = G_+(p-i0) + G_-(p-i0) \, ,
\hspace{0.5cm} p\in [1/\k-y,1/\k-x].
\eeq
Using the parity condition~\rf{parity} this can also be written as
\beq
G_+(p-i0) - G_+(p+i0) = G_-(p-i0) - G_-(p+i0) \, ,
\hspace{0.7cm} p\in [x,y].
\label{nocut}
\eeq
Now, for any complex $p$ not belonging to the intervals $[x,y]$
and $[\frac{1}{\k}-y,\frac{1}{\k}-x]$ we can write $G_+(p)$ as
(cf. figure~\ref{int})
\bea
G_+(p)&=&\oint_{C_0}\frac{d\om}{2 \pi i}\frac{1}{p-\om}G_+(\om)
=\frac{1}{2}\oint_{C_0\cup C_1}\frac{d\om}{2\pi i}\,
\frac{1/\k-2\om}{\left(1/\k-p-\om\right)(p-\om)}
G_+(\om) 
\nonumber \\*
&=&\frac{1}{2}\int_{C_+\cup C_-}\frac{d\om}{2\pi i}
\,\frac{1/\k-2\om}
{\left(1/\k-p-\om\right)(p-\om)}G_+(\om)\non \\*
&=&\oint_{C_-}\frac{d\om}{2 \pi i}\,
\frac{1/\k-2\om}{\left(1/\k-p-\om\right)(p-\om)}
G_+(\om) \nonumber \\*
&=&\oint_{C_-}\frac{d\om}{2\pi i}\,
\frac{1/\k-2\om}{\left(1/\k-p-\om\right)(p-\om)}
G_-(\om)
\label{G+-}
\eea
where the last equality sign follows from~\rf{nocut} and the one
before that from the parity condition~\rf{parity}. Hence,
 once we know 
$G_-(p)$ we can determine $G_+(p)$. Adding the two eliminates the cut
$[\frac{1}{\k}-y,\frac{1}{\k}-x]$ and produces the full transfer
matrix $G(p)$.
Let us mention that by arguments 
analogous to those above one can also express $G_-(p)$ via a
contour integral involving $G_+(p)$, namely
\beq
G_-(p)=\oint_{C_-}\frac{d\om}{2\pi i}\,
\frac{1/\k-2p}{\left(1/\k-p-\om\right)(p-\om)}G_+(\om).
\label{G-+}
\eeq
In order to determine the inverse Laplace transform of 
$G(p)$ i.e.\ in order to determine the generating function for the
loop-loop amplitudes it is sufficient to know $G_-(p)$; namely
\beq
G(l)=\oint_{C_-}\frac{dp}{2\pi i}\,p^l G_-(p)
\label{Gl}
\eeq
which follows immediately from the definition~\rf{defG+-}. 
We note that the relations~\rf{G+-}, \rf{G-+} and~\rf{Gl} 
are true for any function
with an analyticity structure similar to that of $G(p)$.

\newsection{The $c=-2$ string \label{c=-2}}

As already mentioned the discrete version of the $c=-2$ string can by
means of Parisi-Sourlas dimensional reduction be mapped onto a
zero-dimensional field theory which can be viewed as a special version
of the $O(n)$ model on a random lattice~\cite{Dav85}. 
More precisely one has to take $n=-2$ and the potential 
to be of the form 
\beq
V(p)=\frac{1}{2}(p-\k p^2)^2 \, ,
\label{V(p)} 
\eeq
which means that 
$g_3 = 3 \k$, $g_4 = -2 \k^2$, and otherwise $g_i=0$. 
A continuum limit can be defined when $\k$ approaches one of
its critical
values \mbox{$\k_c=\pm\frac{1}{8}$}. We note that the potential has exactly the
properties assumed in equation~\rf{V'} and~\rf{g5=0}. 

\subsection{The disk amplitude}
The disk amplitude $W_-(p)$
which enters the differential
equation~\rf{dTf} can be found in several
references, see for instance~\cite{Dav85,KS92,EK95}. It can also be derived 
entirely within the string field theory framework generalising the ideas of
reference~\cite{Wat95}. One has
\beq
W_-(p) - \frac{1}{2} V'(p) = 
- \k^2(p-\frac{1}{2\k})\sqrt{(p-x)(p-y)(p-1/\k+x)(p-1/\k+y)}
\label{oneloop}
\eeq
where 
\beq
x=\frac{1}{2\k}
-\left(\frac{1}{4\k^2}+\frac{2}{\k}\right)^{1/2} 
\label{x}, \hspace{1.0cm}
y=\frac{1}{2\k}
-\left(\frac{1}{4\k^2}-\frac{2}{\k}\right)^{1/2}.
\label{xandy}
\eeq
We shall be interested in studying the model 
in the vicinity of the critical point
$\k_c=\frac{1}{8}$. (We choose to consider $\kappa_c>0$ since this gives
us positive values for the amplitudes.)
In this region of the parameter space we have 
$x<y\leq\frac{1}{2\k}\leq\frac{1}{\k}-y<\frac{1}{\k}-x$. 
The square root in~\rf{oneloop} is defined so that
its cuts are $[x,y]$ and $[1/\k-y,1/\k-x]$ and so that it is positive
as $p\rightarrow \infty$. 
In particular this implies that the square root is negative 
for $p \in\,]y,1/{\kappa}-y[$. We note that the cut structure
of~\rf{oneloop} is exactly as depicted in figure~\ref{int} and that
the critical point corresponds to the situation 
where the two cuts merge.

Let us now proceed to taking the continuum limit. To do so we must
scale the coupling constant, $\k$, which is the discrete analogue of 
the cosmological constant, towards its critical value. We choose the
following prescription
\beq
\k = \k_c e^{- \epsilon^2 \Lambda}
\label{kscal}
\eeq 
where $\Lambda$ is the continuum cosmological constant and $\epsilon^2$
is a scaling parameter with the dimension of volume. Given the scaling of
$\k$ the boundary equations~\rf{xandy} tell us how $x$ and $y$ will
behave in the scaling limit. We have
\beq
y = \frac{1}{2\k_c}( 1-\epsilon\,\sqrt{\Lambda}+{\cal O}(\epsilon^3) ),
\hspace{1.0cm}
x = \frac{1}{2\kappa_c}( 1-\sqrt{2}+{\cal O}(\epsilon^2) ).
\eeq
In order to get a
non-trivial scaling of the one-loop correlator we must likewise scale 
$p$ to $\frac{1}{2\k_c}$ and this must be done in such a way that $p$ 
always remains inside the region of convergence of $W_-(p)$. We set 
\beq
p = p_c e^{\epsilon \, \sigma} , 
~~~~
p_c = \frac{1}{2\k_c} ,
\label{pscal}
\eeq
where we note that in case $\sigma$ is real it must belong to the
interval $[-\sqrt{\Lambda},\sqrt{\Lambda}]$. 
Then we find for the one-loop correlator, 
\beq
W_-(p) - \frac{1}{2} V'(p) = 
\epsilon^2 \sqrt{2}\,\sigma\sqrt{\Lambda-\sigma^2}
+ {\cal O}(\epsilon^3) 
\label{W-sig0}
\eeq
and we define its continuum version, $W_-(\sigma)$, by
\beq
W_-(\sigma) = \lim_{\epsilon \rightarrow 0} 
\frac{1}{\sqrt{2} \epsilon^2} \left( W_-(p) - \frac{1}{2} V'(p) \right) 
= \sigma\sqrt{\Lambda-\sigma^2} .
\label{W-sig}
\eeq
Here the square root is defined so that it has two cuts,
$[-\infty,-\sqrt{\Lambda}]$ and $[\sqrt{\Lambda},\infty]$ and so that
it is positive for $\sigma\in \, ]\!-\sqrt{\Lambda},\sqrt{\Lambda}[$. 
This structure is dictated by the original definition of 
the square root in~\rf{oneloop}. 
Let us note that while $W_-(p)$ and $W_+(p)$, whether
in their discrete or continuum version, both have two cuts, the full
one-loop correlator of course has only one cut. 
Exactly as for the transfer matrix there exist contour integral 
formulas which connect $W_+(p)$ and $W_-(p)$ 
(cf.\ equation~\rf{G+-} and~\rf{G-+}). 
%The result is 
%\beq\label{AmpContinuum}
%W(\sigma) = 
%- \frac{\sigma}{2\sqrt{2}\epsilon} 
%- \frac{\sigma^2 - 2\Lambda}{4\sqrt{2}} 
%+ \frac{2}{\pi} \sigma \sqrt{\Lambda - \sigma^2} 
%\arctan \sqrt{ \frac{\sqrt{\Lambda}-\sigma}{\sqrt{\Lambda}+\sigma} } \, .
%\eeq
%Also the recipe~\rf{Gl} for finding the inverse Laplace transform 
%applies to $W(p)$.
%We thus find 
%\beq\label{AmpContinuum2}
%W(L) 
%= - \int_{-\infty}^{-\sqrt{\Lambda}} \frac{d\sigma}{\pi}\,
%  e^{\sigma L}\,\sigma\sqrt{\sigma^2-\Lambda}
%= \frac{\Lambda}{\pi L}K_2(\sqrt{\Lambda}L) \, , 
%\eeq
%which agrees with the result of references~\cite{MSS91,EK91}. 
%The inverse Laplace transformation of \rf{AmpContinuum2} 
%with respect to $\Lambda$ gives 
%\beq\label{AmpContinuum3}
%W_V(L) = \frac{L}{8 \pi V^3} e^{- L^2 /(4V) } \, . 
%\eeq
%{}From \rf{AmpContinuum3} 
%We expect that 
%$W_V(L) \sim L\,V\,Z_V$ for $L \sim 0$, 
%where $Z_V$ is the partition function for universes with   volume $V$. 
%Then, from~\rf{AmpContinuum3}
% we find that $Z_V = V^{-4}(= V^{\gamma_{\rm str}-3})$, 
i.e.\ $\gamma_{\rm str} = -1$. 

\subsection{The transfer matrix \label{transfer}}

We shall now explicitly determine the transfer matrix for the $c=-2$
string by solving the differential equation~\rf{dTf}. For simplicity
we will work in the continuum language. In the previous section we
have shown how to take the continuum limit of the one-loop
correlator. Let us now discuss how to take the continuum limit of the
remaining terms in equation~\rf{dTf}. As regards the arguments of
$G_-(p,q,t)$, obviously the scaling of $p$ is dictated by the
relation~\rf{pscal}. The scaling of $q$, on the other hand, is
determined by the fact that the transfer matrix must obey a simple 
composition law~\cite{KKMW93} and reads
\beq
q = q_c e^{\epsilon \, \tau} \, ,
~~~~
q_c = \frac{1}{p_c} = 2 \k_c \, . 
\label{scalq}
\eeq 
The necessary scaling of $t$, the distance or time evolution
parameter, then follows from the structure of the differential
equation. We see that $t$ must behave as $t\sim\epsilon^{-1}\, T$ in order
for equation~\rf{dTf} to be consistent and we set
\beq
t = \frac{p_c}{2\sqrt{2} \epsilon} \, T.
\label{tscal}
\eeq
In particular we see that $T$ has the dimension of (volume)$^{1/2}$.
Inserting the scaling
relations~\rf{pscal} and~\rf{scalq} into the initial 
condition~\rf{diffbound}
we see that it is natural to introduce a continuum version of the
transfer matrix by
\beq
G(p,q,t)\rightarrow \frac{1}{\epsilon}G(\sigma,\tau,T)
\eeq
and from~\rf{pscal} it follows that the parity condition~\rf{parity} turns
into
\beq
G_{\pm}(\sigma,\tau,T)
=\pm G_{\pm}(-\sigma,\tau,T).
\eeq
Collecting everything we can write our differential equation as
\bea
\frac{\partial}{\partial T}G_-(\sigma,\tau,T)
&=&-\frac{\partial}{\partial \sigma}\left(
W_-(\sigma)\,G_-(\sigma,\tau,T)\right), \label{dTff}
\\*
G_-(\sigma,\tau,T=0)&=&\frac{2\sigma}{\sigma^2-\tau^2}
\eea
where $\sigma$ has to belong to the region of analyticity of the square
root, i.e.\ $\sigma$ must lie outside the intervals 
$[-\infty,-\sqrt{\Lambda}]$ and $[\sqrt{\Lambda},\infty]$. The procedure
for solving an equation like~\rf{dTff} is standard. The solution is
given by
\beq
G_-(\sigma,\tau,T)
= \frac{W_-(\hat\sigma)}{W_-(\sigma)} \, G_-(\hat\sigma,\tau,T=0) 
= \frac{\hat\sigma\,\sqrt{\Lambda-\hat{\sigma}^2}}
       {\sigma\,\sqrt{\Lambda-\sigma^2}}\, 
  \frac{2\hat\sigma}{\hat{\sigma}^2-\tau^2}
\label{G-}
\eeq
where $\hat{\sigma}=\hat{\sigma}(\sigma,T)$ 
is a solution of the characteristic equation
\beq
\frac{\partial}{\partial T}\,\hat{\sigma}(\sigma,T) 
= -\hat{\sigma} \sqrt{\Lambda-\hat{\sigma}^2}
\label{character}
\eeq
with boundary condition
\beq
\hat{\sigma}(\sigma,T=0)=\sigma. 
\label{charbound}
\eeq
The equations~\rf{character} and~\rf{charbound} imply
\beq
T = -\int_{\sigma}^{\hat{\sigma}(\sigma,T)}
     \frac{d\sigma'}{ \sigma'\sqrt{\Lambda-\sigma'^2}}
  = \left[\frac{1}{ \sqrt{\Lambda}}
    \log\left(\frac{\sqrt{\Lambda}+\sqrt{\Lambda-\sigma'^2}}{\sigma'}\right)
    \right]^{\hat\sigma(\sigma,T)}_{\sigma} \, .
\label{charint}
\eeq
{}From the integral in~\rf{charint} 
we conclude that in case $\sigma$ is real (and hence belongs to the
interval $[-\sqrt{\Lambda},\sqrt{\Lambda}]$) we have $\hat{\sigma}(T)<
\hat{\sigma}(0)=\sigma$. 
{}From~\rf{charint} we get
%\beq
%e^{2\sqrt{\Lambda}T}
%= \frac{ \Lambda+\sqrt{\Lambda-\hat{\sigma}^2} }
%       { \Lambda+\sqrt{\Lambda-\sigma^2} }\,\frac{\sigma}{\hat\sigma}.
%\label{expD}
%\eeq
\beq
\hat{\sigma}(\sigma,T) = 
\frac{\sigma}{\cosh(\sqrt{\Lambda}T) +
\sqrt{1-\frac{\sigma^2}{\Lambda}}\sinh(\sqrt{\Lambda} T)} \, . 
\label{sigmahat}
\eeq
{}From this equation we see that $\hat{\sigma}(\sigma,T)$ is an odd function of
$\sigma$ and that $\hat{\sigma}(\sigma,T)\rightarrow 0$ as
$T\rightarrow \infty$. 
We can also write~\rf{sigmahat} as
\beq
\sqrt{\Lambda-\hat{\sigma}^2}=
\sqrt{\Lambda}\left\{
\frac{\sqrt{\Lambda}\sinh(\sqrt{\Lambda}T)+
\sqrt{\Lambda-\sigma^2}\cosh(\sqrt{\Lambda}T)}
{\sqrt{\Lambda}\cosh(\sqrt{\Lambda} T)+
\sqrt{\Lambda-\sigma^2}\sinh(\sqrt{\Lambda} T)}
\right\}.
\label{sqrtsig}
\eeq
This relation will prove very convenient for our later
considerations. 
{}From equations~\rf{G-}, \rf{sigmahat} and~\rf{sqrtsig}
we see that $G_-(\sigma,\tau,T)$ has the properties that we
expect. First, it is an odd function of $\sigma$. 
Furthermore, we see that it is analytical in $\sigma$ except for two cuts,
$[-\infty,-\sqrt{\Lambda}]$ and $[\sqrt{\Lambda},\infty]$. 
In particular, we note that it has no poles in $\sigma$. 
This means that
we can indeed use the formulas derived in section~\ref{n=-2} to pass from
$G_-(\sigma)$ to $G_+(\sigma)$ and vice versa. 
The continuum version of
the relations~\rf{G+-} and~\rf{G-+} read
\bea
G_+(\sigma)&=&\int_{-\infty}^{-\sqrt{\Lambda}}\frac{d\om}{2 \pi i}\,
\frac{2\om}{\sigma^2-\om^2}\left\{G_-(\om-i0)-G_-(\om+i0)\right\},
\label{G+-cont} \\*
G_-(\sigma)&=&\int_{-\infty}^{-\sqrt{\Lambda}}\frac{d\om}{2 \pi i}\,
\frac{2\sigma}{\sigma^2-\om^2}\left\{G_+(\om-i0)-G_+(\om+i0)\right\}
\label{G-+cont}
\eea
which can be seen simply by inserting the scaling relation~\rf{pscal}
for $p$ into~\rf{G+-} and~\rf{G-+}.
Inserting the expression~\rf{G-} for $G_-(\sigma,\tau,T)$ into~\rf{G+-cont}
we get the even part of the transfer matrix which, as the odd part, is analytic
in $\sigma$ except for the two cuts $[-\infty,-\sqrt{\Lambda}]$ and
$[\sqrt{\Lambda},\infty]$. Adding $G_-(\sigma,\tau,T)$ and 
$G_+(\sigma,\tau,T)$ eliminates the unphysical cut $[\sqrt{\Lambda},\infty]$
and produces the full transfer matrix.
In analogy with the discrete case, in order to determine the inverse Laplace
transform of $G(\sigma)$ it is sufficient to know $G_-(\sigma)$. The
continuum version of~\rf{Gl} reads
\beq
G(L) = \int_{-\infty}^{-\sqrt{\Lambda}}\frac{d\sigma}{2\pi i}
e^{\sigma L}\left\{
G_-(\sigma-i0)-G_-(\sigma+i0)\right\}\equiv
\oint_{C}\frac{d\sigma}{2\pi i}e^{\sigma L}G_-(\sigma)
\label{GL}
\eeq
where $L$ is related to $l$ by $L=l \epsilon$. The
relations~\rf{G+-cont}, \rf{G-+cont} and~\rf{GL} are valid for any
function with an analyticity structure similar to that of
$G(\sigma)$. In
particular they hold also for $W(\sigma)$.

\subsection{The two-point function}
\begin{figure}[htb]
\centerline{\epsfbox{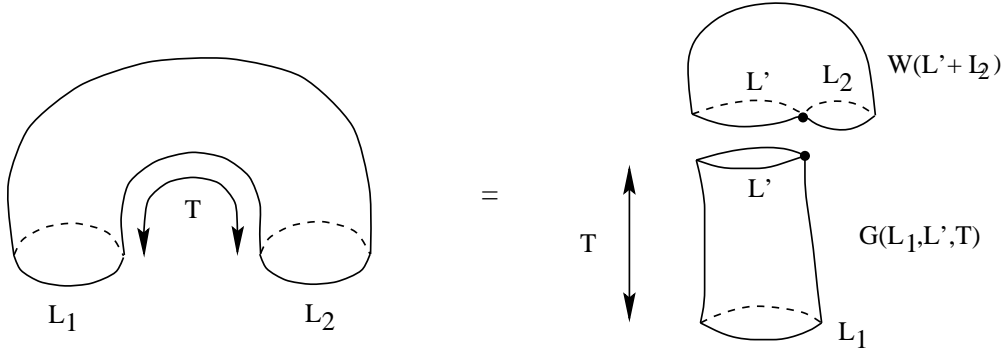}}
%\centerline{\psfig{figure=twopointc.ps,width=14cm,angle=-90}}
\caption{The construction of the two-point function from the loop-loop
correlator and the disk amplitude. (The loop lengths $L_1$ and $L_2$ 
are sent to zero).}
\label{twopoint} 
\end{figure}

In order to calculate the two-point function we shall use the strategy of
reference~\cite{AW95,nishi} of expressing it in terms of the loop-loop
correlator and the disk amplitude. 
We remind the reader that the transfer matrix is nothing but the
generating functional for loop-loop correlators.
In the present case we
shall work directly in the scaling limit. 
Using continuum notation we write the two-point function 
as (cf.\ figure~\ref{twopoint})
\beq
G_{\Lambda}(T) = 
\frac{-1}{\log \epsilon}
\lim_{L_1\rightarrow \epsilon}
\lim_{L_2\rightarrow \epsilon}
{\cal G}_{\Lambda}(L_1,L_2,T)
\label{recipe}
\eeq
where 
\beq
{\cal G}_{\Lambda}(L_1,L_2,T) = \frac{1}{L_1} 
\int_{0}^{\infty}\!dL' \, G(L_1,L',T)\,L'\,W(L'+L_2) \, .
\label{GD}
\eeq
Here $W(L)$ is the
continuum disk amplitude related to the continuum one-loop correlator
$W(\sigma)$ by inverse Laplace transformation. 
Similarly
$G(L_1,L_2,T)$ is related to $G(\sigma,\tau,T)$ by inverse Laplace
transformation in $L_1$ and $L_2$. 
The factor $L'$ introduces a marked point on the exit loop of the 
cylinder amplitude while the factor $1/L_1$ removes that on the entrance loop.
The object $G_{\Lambda}(L_1,L_2,T)$ does not have marked points on either
its entrance or its exit loops and
is invariant %symmetric 
under the exchange of $L_1$ and $L_2$, 
i.e.\ 
${\cal G}_{\Lambda}(L_1,L_2,T) = {\cal G}_{\Lambda}(L_2,L_1,T)$,
(cf.\ figure~\ref{twopoint}).
%The presence of the factor $L'$ in the integral~\rf{GD} 
%is due to our convention of introducing a marked
%link only on the entrance loop of the cylinder amplitude. 
%This means that 
%there are $L'$ different ways of gluing together the two objects
%appearing in the integrand of equation~\rf{GD} 
A cut-off, $\epsilon$,  has been introduced in order to regularise the 
divergence for small $L_1$ and $L_2$. 
The division by the factor $\log\epsilon$ is 
likewise introduced for regularisation purposes. 
We expect this divergence because 
%as we shall see later, 
the disk amplitude $W(L)/L$ behaves as $\log L$ for small $L$. 
Introducing explicitly the inverse
Laplace transform of the transfer matrix we get
\bea
\lefteqn{\hspace{-12pt}
{\cal G}_{\Lambda}(L_1,L_2,T) =
\frac{1}{L_1} \int_{0}^{\infty}\!dL' 
\int_{-i\infty}^{i\infty} \frac{d\sigma}{2\pi i}\,e^{\sigma L_1}\!
\int_{-i\infty}^{i\infty} \frac{d\tau}{2\pi i}\,e^{\tau L'}\,
G_-(\sigma,\tau,T) L' \, W(L'+L_2) }
\label{sG-}\\
&&\hspace{20pt} =  
\frac{1}{L_1} \int_{-i\infty}^{i \infty} \frac{d\sigma}{2\pi i}\, e^{\sigma L_1}\!
\int_{-i\infty}^{i\infty}\frac{d\tau}{2\pi i}\,
\frac{\hat{\sigma}\sqrt{\Lambda-\hat{\sigma}^2}}
{\sigma\sqrt{\Lambda-\sigma^2}}\frac{2\hat{\sigma}}{\hat{\sigma}^2-\tau^2} 
\int_{0}^{\infty}\!dL' 
e^{\tau L'} L' \, W(L'+L_2) \, . 
\nonumber
\eea
Picking up the residue at $\tau=-\hat{\sigma}$ we get
\bea
{\cal G}_{\Lambda}(L_1,L_2,T) 
&=& 
\frac{1}{L_1} \int_{-i \infty}^{i\infty}\frac{d\sigma}{2\pi i}\,e^{\sigma L_1}\!
\frac{\hat{\sigma}\sqrt{\Lambda-\hat{\sigma}^2}}{
      \sigma\sqrt{\Lambda-\sigma^2}}
\int_{0}^{\infty}\!dL'\, e^{-\hat{\sigma}L'} L' \, W(L'+L_2) 
\nonumber \\
&=& 
- \frac{1}{L_1} \int_{-i\infty}^{i\infty}\frac{d\sigma}{2\pi i}\,e^{\sigma L_1}\!
\frac{\hat{\sigma}\sqrt{\Lambda-\hat{\sigma}^2}}{
      \sigma\sqrt{\Lambda-\sigma^2}}
\frac{\partial}{\partial \hat{\sigma}} W_{L_2}(\hat{\sigma}) 
\, , 
\label{sG-b}
\eea
where $W_L(\sigma)$ is a kind of cylinder amplitude defined by 
\beq\label{Cylinder}
W_L(\sigma) = 
\int_{0}^{\infty}\!dL' \, e^{-\sigma L'} \, W(L'+L) \, .
\eeq
Using the relation 
$\partial \hat{\sigma} / \partial \sigma = 
 \hat{\sigma} \sqrt{\Lambda-\hat{\sigma}^2} / \sigma \sqrt{\Lambda-\sigma^2}$,
we find 
\beq\label{univ}
{\cal G}_{\Lambda}(L_1,L_2,T) = 
- \frac{1}{L_1} \int_{-i\infty}^{i\infty}\frac{d\sigma}{2\pi i}
\,e^{\sigma L_1}\,
\frac{\partial}{\partial \sigma} W_{L_2}(\hat{\sigma}) .
\eeq
The formula~\rf{univ} has some flavour of universality 
and it does indeed take the same form in the pure gravity case 
(except for $\hat{\sigma}(\sigma,T)$ being differently defined). 
We next determine $W(L)$ and $W_L(\sigma)$. 
Applying the recipe~\rf{GL} we get %up to a factor of $\pi$
\beq
W(L) = - \int_{-\infty}^{-\sqrt{\Lambda}} \frac{d\sigma}{\pi}\,
         e^{\sigma L} \sigma\sqrt{\sigma^2-\Lambda}
     = \frac{\Lambda}{\pi L}K_2(\sqrt{\Lambda}L)
\eeq
which agrees with the result of references~\cite{MSS91,EK91}. 
Then, we find 
\bea
&&W_L(\sigma) = 
\frac{1}{\pi} \, \biggl\{ \, 
\frac{1}{L^2} - \frac{\sigma}{L} 
- \Bigl( \sigma^2 - \frac{\Lambda}{2} \Bigr) 
  \Bigl( \log \frac{\sqrt{\Lambda}L}{2}  +\gamma \Bigr) 
- \frac{\Lambda}{4} 
\label{Cylinder2}\\
&&\hspace{70pt}
+ \, 2 \sigma \sqrt{ \Lambda - \sigma^2 } \, 
\arctan \sqrt{ \frac{\sqrt{\Lambda} - \sigma}{\sqrt{\Lambda} + \sigma} }
\, + \, O(L) 
\, \biggr\}
\, ,
\nonumber
\eea
where $\gamma$ is the Euler constant.
Substituting \rf{Cylinder2} into \rf{sG-b} 
we find 
\bea
{\cal G}_{\Lambda}(L_1,L_2,T)
&=&
\frac{1}{\pi L_1} \int_{-i \infty}^{i\infty}
\frac{d\sigma}{2\pi i}\, e^{\sigma L_1}
\frac{\hat{\sigma}\sqrt{\Lambda-\hat{\sigma}^2}}{
      \sigma\sqrt{\Lambda-\sigma^2}}
%\frac{\partial}{\partial \hat{\sigma}}
\Biggl\{ \, 
\frac{1}{L_2} + 2 \hat{\sigma} 
\biggl( \log\frac{\sqrt{\Lambda}L_2}{2} + \gamma + \frac{1}{2} \biggr)
\nonumber \\ 
&&\hspace{42pt}
+ \, \frac{2(2\hat{\sigma}^2-\Lambda)}{\sqrt{\Lambda-\hat{\sigma}^2}} \,
\arctan \sqrt{ \frac{ \sqrt{\Lambda} - \hat{\sigma} }{
                      \sqrt{\Lambda} + \hat{\sigma} }  }
\, + \, O(L_2) \, \Biggr\} .
\eea
By making use of the relations~\rf{sigmahat} and~\rf{sqrtsig} 
it is now possible to carry out explicitly 
the $\sigma$-integration to
the leading order in $L_1$ and $L_2$. 
We get (up to a factor of $4/\pi^2$)
\bea
&&{\cal G}_{\Lambda}(L_1,L_2,T) = 
\frac{\Lambda^{3/2}}{2}\,\Biggl\{ 
  \frac{\cosh(\sqrt{\Lambda}T)}{\sinh^3(\sqrt{\Lambda}T)}
  \biggl( \log\frac{\sqrt{\Lambda}L_1}{2}  \biggr)
  \biggl( \log\frac{\sqrt{\Lambda}L_2}{2} \biggr)
\label{GL1T} \\
&&\hspace{48pt}
- \, \frac{1+\cosh^2(\sqrt{\Lambda}T)}{4\sinh^3(\sqrt{\Lambda}T)}
  \log\biggl(\frac{\cosh(\sqrt{\Lambda}T)-1}{\cosh(\sqrt{\Lambda}T)+1}\biggr) 
  \biggl( \log\frac{\Lambda L_1 L_2}{4}  \biggr)
+ \ldots\ldots 
\Biggr\} \, . 
\nonumber
\eea
Applying the recipe~\rf{recipe} to get $G_{\Lambda}(T)$ we find
\bea
G_{\Lambda}(T) 
&=& 
\frac{\Lambda^{3/2}}{\sinh^3(\sqrt{\Lambda}T)}
\left\{ - \cosh(\sqrt{\Lambda}T) 
\Bigl( \log\frac{\sqrt{\Lambda}\epsilon}{2} \Bigr)
\right.
\nonumber \\
&&\hspace{1.0cm}
\left. 
+ \, \frac{1}{4} \Bigl( 1 + \cosh^2(\sqrt{\Lambda}T) \Bigr) 
\log\biggl(\frac{\cosh(\sqrt{\Lambda}T)-1}{\cosh(\sqrt{\Lambda}T)+1}\biggr) 
\right\}
\label{GLf}
\eea
where we have left out sub-leading terms 
and terms expandable in integer powers of $\Lambda$. At first sight it might seem that the second term
in~\rf{GLf} (originating from the third line of~\rf{GL1T}) is also
sub-dominant. However, this is only true for finite $T$. When $T$ is of
the same order as $\epsilon$, the two terms in~\rf{GLf}
are of the same order of magnitude.
 As we shall see in the following section it is 
of utmost importance that both terms are taken into account.

\subsection{Critical properties and fractal structure}

The critical indices 
$\nu$ and $\eta$  are defined by (see ref.\ \cite{AW95})
\beq
G_\Lambda(T) \ \sim \ 
  \left\{
    \begin{array}{ll}
       e^{-({\rm const.})\Lambda^\nu T}  & \qquad
        \mbox{ for~~~~ $T \rightarrow \infty$} \\
       T^{1-\eta}                         & \qquad
\mbox{ for~~~~ $T \rightarrow 0$} 
    \end{array}
  \right. \,   
\eeq
{}From the expression~\rf{GLf} for the two-point function we can
immediately read off the critical index $\nu$.
 Letting $T\rightarrow \infty$ we find
\beq
G_{\Lambda}(T) \sim 
-4\, \Lambda^{3/2} e^{-2\sqrt{\Lambda}T} \log\frac{\sqrt{\Lambda}\epsilon}{2} 
\eeq
from which we conclude that
\beq
\nu = \frac{1}{2}.
\eeq
For small $T$, we can
rewrite the two-point function as
\bea
G_{\Lambda}(T) &\approx& 
\frac{\Lambda^{3/2}}{2\sinh^3(\sqrt{\Lambda}T)} \Biggl\{
  \Bigl( \cosh(\sqrt{\Lambda}T)-1 \Bigr)^2 
  \log\frac{\sqrt{\Lambda}T}{2}
+ 2 \cosh(\sqrt{\Lambda}T)
  \Bigl( \log\frac{T}{\epsilon} \Bigr) 
\nonumber \\
&&\hspace{81pt}
- \, \frac{1}{12}\,\Lambda T^2\, \Bigl( 1+\cosh^2(\sqrt{\Lambda}T) \Bigr) 
\Biggr\} .
\label{G2}
\eea
{}From the expression~\rf{G2} we see that for $\epsilon \ll T \ll \Lambda^{-1/2}$
we have
\beq
G_{\Lambda}(T) \sim 
 \log\left(\frac{T}{\epsilon}\right) \frac{1}{T^3}
\eeq
which means that
\beq
\eta=4.
\eeq
It is well known that $\gamma_{\rm str}=-1$ for the $c=-2$
string~\cite{Dav85,KKM85} and it also appears immediately from either the
expression~\rf{GLf} or~\rf{G2} since the string susceptibility can be calculated
as
\beq
\chi(\Lambda) 
= \int_{{\rm (const.)}\epsilon}^{\infty}\!dT \, G_{\Lambda}(T) 
= \frac{\Lambda}{4} \log \frac{\sqrt{\Lambda}\epsilon}{2} 
\eeq
where we have left out sub-leading singularities and terms analytic in
$\Lambda$. Consequently we have that the quantum gravity version of
Fisher's scaling relation is fulfilled:
\beq
\gamma_{\rm str} = \nu(2-\eta).
\eeq
We note that had we not taken into account the second term in
equation~\rf{GLf} we would not have got the correct string susceptibility.
If we treat $T$ as a measure of geodesic distance we get for the grand
canonical Hausdorff dimension of our manifolds
\beq
d_H\equiv\frac{1}{\nu} = 2 = - \, \frac{2}{\gamma_{\rm str}}.
\eeq

\newsection{Discussion \label{discuss} }
The (internal) Hausdorff dimension of the universe with $c=-2$ matter present
agrees with what one would expect from the dimensional analysis carried out
in section~\rf{transfer} showing that the dimension of $T$ is the same
as that of $V^{1/2}$. By the same type of dimensional analysis we can
predict the (internal) Hausdorff dimension of the space time manifold in the 
case
where a minimal unitary model is coupled to gravity. A minimal
model of the type $(m,m+1)$ can be reached starting from
the $O(n)$ model with $n=2\cos(\pi/m)$ by a suitable fine tuning of the
coupling constants. Only a potential of cubic order is needed and
hence all the equations of section~\ref{O(n)} remain true. For the
$(m,m+1)$ minimal unitary model it is well
known~\cite{KS92,EZ92,EK95} that when we let the coupling constants
$g_3$ and $\kappa$ approach their critical values $g_3^c$ and $\kappa_c$
as
\beq
g_3 - g_3^c \sim - \epsilon^2 \Lambda,\hspace{1.0cm}
\kappa-\kappa_c\sim |g_3-g_3^c|^{1/2}
\label{gunscal}
\eeq
the variable $p$, conjugate to the loop length $l$, will behave as
\beq
p-p_c =p-\frac{1}{2\kappa_c}\sim \epsilon \sigma
\label{punscal}
\eeq
and the singular part of the disk amplitude exhibits the following
scaling
\beq
W_{s}(p)\sim \epsilon^{1+\frac{1}{m}}W_{s}(\sigma),\hspace{1.0cm}
W_{s}\left(\frac{1}{\kappa}-p\right)\sim \epsilon^{1+\frac{1}{m}}
W_{s}(-\sigma).
\eeq
In order for the differential equation~\rf{dT5} 
to be consistent we must hence
require that
\beq
t \sim \epsilon^{-\frac{1}{m}}T.
\label{tunscal}
\eeq
Combining~\rf{tunscal} and~\rf{gunscal} we now see that $T$ has the
dimension of $V^{1/2m}$ and we reach the following prediction
for the grand canonical (internal) Hausdorff dimension
\beq
d_H \equiv \frac{1}{\nu} = 2m
\eeq
which agrees with the prediction of 
references~\cite{IK94,IIKMNS94,Kos95}.
The dimensional analysis can also determine how to take the continuum 
limit of the string coupling $g_s$.  From \rf{H} one finds 
$g_s \sim \epsilon^{4+\frac{2}{m}}G_s$. 
Recently it has been shown that for $n>2$ the $O(n)$ model on a random
lattice possesses new types of critical points at which $\gamma_{\rm str}$
takes the values $\gamma_{\rm str}=\frac{1}{k}$, 
$k=2,3,\ldots$~\cite{EK962,DK97}.
For these models the scaling argument does not immediately
lead to a meaningful prediction for $d_H$.

The above considerations are based on the assumption that we can treat
$T$ as a measure of the geodesic distance.
However, from figure 2 it is obvious
the minimal step decomposition 
is {\it not} directly related to the geodesic distance. 
Nevertheless, we cannot rule out {\it a priori} that $T$ is effectively 
proportional to the geodesic distance when the functional average is 
performed. This is what happens if we compare various reasonable 
``definitions'' of geodesic distance on triangulated surfaces.
Such distances can differ quite a lot for individual triangulations, but when  
the summation over all triangulations is performed and the scaling limit 
is taken they will be proportional. This has also been a basic assumption
behind the identification performed in string field theory 
between $T$ and the geodesic distance. 
We are now in a position where we for the first time can test this assumption. 
For $c=-2$ one predicts $d_H=2$ by assuming that $T$ is proportional to 
the geodesic distance. 
Since this is not a large (internal) Hausdorff dimension and since 
there exists a very efficient computer algorithm for simulation of 
the $c=-2$ theory \cite{KKSW92}, one can test this prediction quite precisely.
Simulations show quite convincingly that 
$d_H \neq 2$~\cite{KKSW92,AAIJKWY96}. 
The most recent high precision measurement 
of $d_H$ gives the following value \cite{AAIJKWY96}
\begin{equation}
d_H = 3.58\pm 0.04,
\label{haus3}
\end{equation} 
with a conservative error estimate. We notice that this value is 
in accordance with the formula 
\begin{equation}
d_H = 2 \times \frac{\sqrt{25-c}+\sqrt{49-c}}{\sqrt{25-c}+\sqrt{1-c}} 
    = 3.561\ldots ,
\end{equation}
derived from the study of diffusion in quantum Liouville theory
\cite{Kaw92}.

Our conclusion is that the variable $T$,  introduced so far in 
non-critical string field theory, is  very natural from the 
point of view of scaling, but is not necessarily related to 
the geodesic distance.
For $c=0$ it {\it is} the geodesic distance. For $c=-2$ it {\it is not} 
the  geodesic distance and the same conclusion is presumably 
true also for $c>0$, although the numerical evidence is not as decisive
as for $c=-2$. The relation between geodesic distance and the string field
time $T$ is one of few major unsolved questions in two-dimensional 
quantum gravity.

\vspace{12pt}
\noindent
{\bf Acknowledgements}\hspace{0.3cm} C. Kristjansen acknowledges the
support of the JSPS and the hospitality of Tokyo Institute of Technology
and Y. Watabiki acknowledges the support and the hospitality of the 
Niels Bohr Institute.

\end{document}